\documentclass{aip-cp}

\usepackage[numbers]{natbib}
\usepackage{rotating}
\usepackage{graphicx}

\begin{document}

% Title portion
\title{Impact of pion dynamics on nuclear shell structure}

\author[aff1,aff2]{Elena Litvinova\corref{cor1}}
%\eaddress[url]{http://tesla.physics.wmich.edu/Bio/Elena _Litvinova/}
%\author[aff2,aff3]{Author's Name}
%\eaddress{anotherauthor@thisaddress.yyy}

\affil[aff1]{Department of Physics, Western Michigan University,
Kalamazoo, MI 49008-5252, USA}
\affil[aff2]{National
Superconducting Cyclotron Laboratory, Michigan State University,
East Lansing, MI 48824-1321, USA}
%\affil[aff3]{You would list an author's second affiliation here.}
\corresp[cor1]{Corresponding author: elena.litvinova@wmich.edu}

\maketitle

\begin{abstract}
Spin-isospin response in exotic nuclear systems is investigated. It is found that in some nuclei excitations with pionic quantum numbers (0$^-$, 1$^+$, 2$^-$, ...) appear at very low energies with large transition probabilities, which is an indication of the vicinity of the onset of pion condensation. As an example, $2^-$ components of the spin-dipole resonance in $^{78}$Ni and $^{132}$Sn are considered. The existence of such modes points out to the necessity of taking into account their coupling to other elementary modes of excitation, e.g. single-quasiparticle ones. This coupling is introduced in the theory for the first time.
Thereby, pion-exchange contribution to the nucleon-nucleon interaction is included in the relativistic framework beyond the Hartree-Fock approximation. Namely, classes of Feynman diagrams are selected according to their significance for nuclear spectroscopic characteristics, such as single-particle energies and strength functions, and included into the nucleonic self-energy in all orders of meson-exchange. As an illustration, the impact of these new contributions on the single-particle energies of $^{100}$Sn is discussed.\end{abstract}

% Head 1
\section{INTRODUCTION}
Nuclear forces are known to have an isospin dependence, which is revealed in many experimental observations. Theoretically, the isospin dependence is modeled by introducing specific terms into the nucleon-nucleon effective interaction or density functional. The underlying mechanism for these terms is the exchange of mesons which carry non-zero isospin, such as rho-mesons and pions. They are included, for instance, in relativistic models on the Hartree or Hartree-Fock (HF) level \cite{SW.86,Long}, and in chiral effective field theories the pion exchange is 
commonly 
included perturbatively \cite{EHM.09,ME.11}. 

Such processes as beta decay, electron capture or charge-exchange reactions represent sensitive probes of nuclear spin-isospin response and, hence, of spin and isospin dependence of the effective nucleon-nucleon interaction. This type of response has been investigated theoretically over decades within approaches with explicit separation of pionic degrees of freedom, in particular, by the so-called 'Moscow school' \cite{Mig.74,TSMM.75,STF.75,MSMM.76,FST.77,TKC.77,KM.77,FST.79,MSTV.90} and 'Stony Brook school' \cite{BB.73,BW.76,BCDM.75,RGG.76,TW.78,TW.79,Meyer.81,DFMW.83}. It has been found that in medium-mass and heavy nuclei as well as in nuclear matter a considerable softening of the pion modes (states with unnatural parity in both neutral and charge-exchange channels) occurs because of the medium effects. Calculations for finite nuclei proved that states of this type can appear at relatively low excitation energies, at least in closed shell medium-mass and heavy nuclei. 

The prevailing view, established from these calculations, is that densities higher than those typically reached in nuclei are needed for pion condensation and, most probably, the condensation does not occur in nuclei. Moreover, even if pion condensation occurred, no observables were found to be sensitive to the presence of the condensate \cite{MSTV.90} and, therefore, to be able to verify its presence.

Recent theoretical developments, however, offer more opportunities for investigations of the soft modes in the spin-isospin channel.   
Self-consistent proton-neutron relativistic quasiparticle random phase approximation (pn-RQRPA) \cite{PNVR.04,NMVPR.05,Liang}  describes the positions of the giant Gamow-Teller resonance (GTR) and the spin dipole resonance (SDR) very reasonably, while the approach beyond QRPA, which accounts for particle-vibration coupling (PVC) effects in the response function \cite{MLVR.12,LBFMZ.14}, improves the description further by the inclusion of the damping mechanism, which is essential for the description of the widths, fine structure and low-lying strength directly related to beta-decay half-lives.

Based on the latter approaches, this work focuses on some peculiarities of the spin-isospin response in exotic nuclei, in particular, on the enhanced low-lying strength of high multipoles. One of the consequences of this enhancement, namely the strong coupling of the low-lying spin-isospin modes to single-nucleon motion, is investigated numerically  for the case of $^{100}$Sn.  

\section{SPIN-ISOSPIN RESPONSE OF NEUTRON-RICH NUCLEI: LOW-LYING STATES}

Relativistic approaches to spin-isospin response based on the covariant density functional theory (CDFT)
\cite{PNVR.04,NMVPR.05,Liang,MLVR.12,LBFMZ.14} form a very convenient framework for calculations for finite nuclei. Starting from a few CDFT parameters (6 parameters for the NL3 interaction or 8 for the density-dependent meson-exchange (DDME) ones \cite{R.06}), a very good description of the GTR and SDR has been achieved. In the proton-neutron relativistic time blocking approximation (pn-RTBA) \cite{MLVR.12}, the effective spin-isospin interaction has the following form:
\begin{eqnarray}
{\tilde V}(1,2;\omega) =
g_{\rho}^2{\vec\tau}_1{\vec\tau}_2(\beta\gamma^{\mu})_1(\beta\gamma_{\mu})_2
D_{\rho}({\bf r}_1,{\bf r}_2) 
- \Bigl(\frac{f_{\bf\pi}}{m_{\pi}}\Bigr)^{2}{\vec\tau}_1{\vec\tau}_2({\bf\Sigma}_1{\bf\nabla}_1)
({\bf\Sigma}_2{\bf\nabla}_2)D_{\pi}({\bf r}_1,{\bf r}_2) + 
g^{\prime}\Bigl(\frac{f_{\pi}}{m_{\pi}}\Bigr)^2{\vec\tau}_1{\vec\tau}_2{\bf\Sigma}_1{\bf\Sigma}_2
\delta({\bf r}_1 - {\bf r}_2) +\nonumber\\ + \Phi_{PVC}(1,2;\omega) - \Phi_{PVC}(1,2;0) = {\tilde V}_{\rho} + {\tilde V}_{\pi} + {\tilde V}_{\delta\pi} +
W_{PVC},
\end{eqnarray}
while in QRPA theories the last terms with $\omega$-dependence marked with 'PVC', which are responsible for damping effects, are absent.
In the approaches without the Fock term there is an additional parameter $g^{\prime}$ in front of the so-called Landau-Migdal term,  which is adjusted to the GTR position in $^{208}$Pb. A detailed description of the pn-RTBA with a dynamical contribution from PVC can be found in Refs. \cite{MLVR.12,LBFMZ.14}.  An analysis of the diagonal matrix elements of the interaction (${\tilde V}_\rho + {\tilde V}_\pi + {\tilde V}_{\delta \pi}$) of Ref. \cite{LZRRM.12} shows that the contribution from the $\rho$-meson exchange is very small, so that in pn-RRPA the positions of the spin-isospin excitations relative to unperturbed single-quasiparticle ones are associated mainly with pionic contributions.
%
% Figure
%
\begin{figure}[h]
  \centerline{\includegraphics[width=450pt]{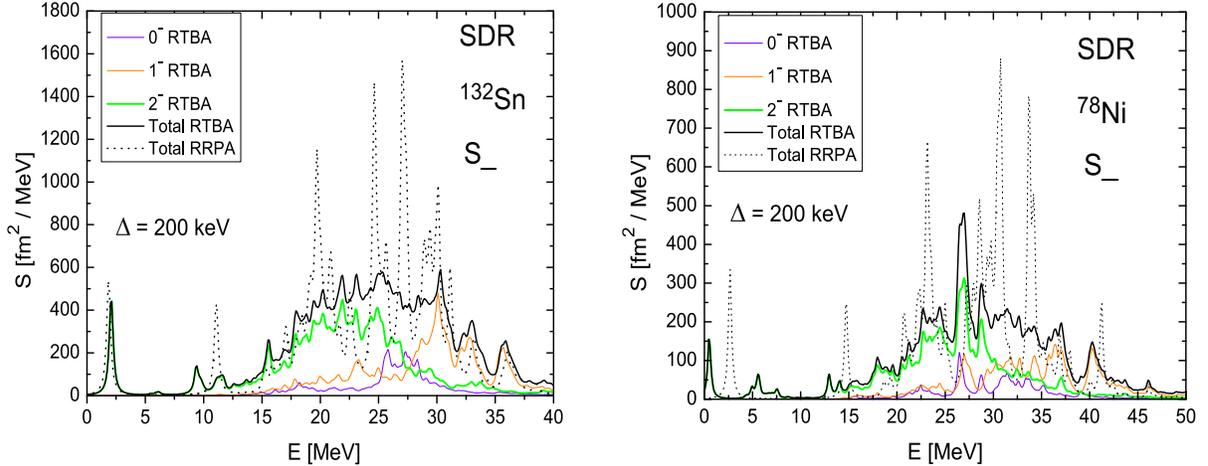}}
  \caption{Spin dipole resonance in neutron-rich nuclei $^{132}$Sn and $^{78}$Ni calculated in pn-RRPA and pn-RTBA. 0$^-$, 1$^-$ and 2$^-$ components computed in pn-RTBA are shown by different colors. The dashed curve displays the total pn-RRPA strength which is a sum of the three components. The black solid curve shows the total pn-RTBA strength.}
\label{sdr} 
\end{figure}

SDR is a nuclear response to the spin-isospin flip operator containing, in addition, the transfer of one unit of angular momentum, i.e. to the operator:
\begin{equation}
P_{\pm}^{\lambda} = \sum\limits_{i=1}^{A} r(i)[\sigma (i)\otimes Y_1(i)]^{\lambda}\tau_{\pm}(i),
\label{psdr}
\end{equation}
with $\lambda = 0,1,2$ and parity $\pi = -1$. Components with different values of $\lambda$ can sometimes be disentangled in charge-exchange reaction experiments on stable nuclei, however, in some cases only the total strength is available. The first calculations of SDR within pn-RTBA were performed for $^{90}$Zr and $^{208}$Pb in Ref. \cite{MLVR.12} and demonstrated a good agreement to data after the inclusion of PVC effects. In both nuclei the proton-neutron SDR shows up in the spectrum as a broad bump localized between 20 and 40 MeV without noticeable structure peculiarities, except a relatively weak soft mode in the $2^-$ channel at about 5 MeV in $^{208}$Pb.

Figure \ref{sdr} displays the (p,n) branch of the spin-dipole resonance, which corresponds to the choice of $\tau_-$ in Eq. (\ref{psdr}), in the neutron-rich nuclei $^{132}$Sn and $^{78}$Ni. The energy axis is related to the ground states of the daughter nuclei $^{132}$Sb and $^{78}$Cu, respectively. Similarly to the cases of $^{90}$Zr and $^{208}$Pb considered previously, the SDR is represented by a broad bump localized mainly between
15-35 MeV ($^{132}$Sn) and 20-40 MeV ($^{78}$Ni). However, one can see strong $2^-$ states at very low energies, whose intensities are comparable to those of the giant resonance.  
Besides the SDR components, we have calculated also strength distributions for higher multipoles with both natural and unnatural parities: $2^{\pm}, 3^{\pm}, 4^{\pm}, 5^{\pm}, 6^{\pm}$. In all channels there are states at very low energies, what shows a competition between various spin-parity pairs for forming the ground-state spins and parities of the odd-odd daughter nuclei. For instance, the lowest pn-RRPA state in $^{78}$Ni  is $4^-$ while there are also $5^-$ and $6^-$ states with very close energies, which is consistent with experimental observations and shell-model calculations for the ground state of $^{78}$Cu. However, it turns out, that in the pn-RRPA approach, which shows a strong fragmentation of SDR, the lowest state in the spectrum can be shifted down as it happens in the case of the $2^-$ state in $^{78}$Ni. This means that PVC can intervene in the competition mentioned above and predict alternative spin-parities of the ground states of odd-odd nuclei, compared to pn-RRPA. Analogously, the calculated spectra of $^{132}$Sn can be related to the states of $^{132}$Sb.

\section{SINGLE-PARTICLE STATES}

\begin{figure}[h]
  \centerline{\includegraphics[width=350pt]{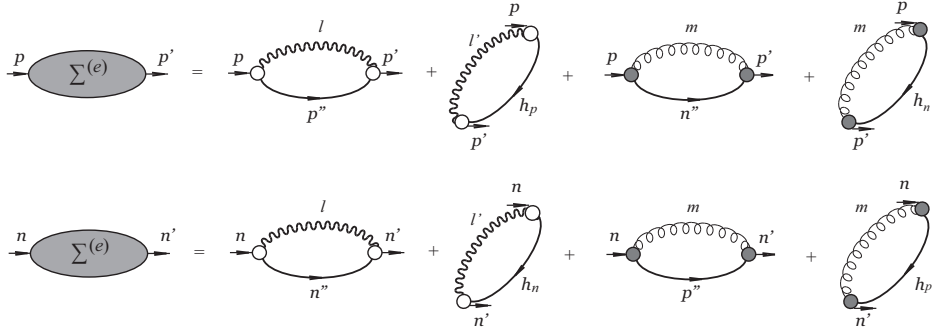}}
  \caption{Isospin structure of the proper self-energy in the second-order of PVC.  The contributions from isoscalar vibrations are represented by the first two terms on the right-hand side (incoming, outgoing and intermediate single-particle states are of the same isospin). The last two terms show the contributions from isospin-flip vibrations (the isospin of the intermediate states is different from the one of the incoming and outgoing states).}
\label{sfe}
\end{figure}

We have seen in the previous section that in spin-isospin channels there are states at very low energies, which can be excited with high probability. A similar situation occurs in the neutral channel where, in particular, very low-lying collective quadrupole and octupole states are present in almost all nuclei throughout the entire nuclear chart. It is known since Bohr and Mottelson \cite{BM.75} that these modes of excitation are very likely to couple to single-particle excitations as well as to other collective modes. Therefore, such a coupling should include also spin-isospin vibrational states which, in analogy to the neutral-channel case, can be called isospin-flip or isovector phonons. If the latter ones are included in the theory, the lowest-order non-local part of the nucleonic self-energy can be expressed diagrammatically as shown in Figure \ref{sfe}.
\vspace{-6cm}
\begin{figure}[h]
  \centerline{\includegraphics[width=300pt]{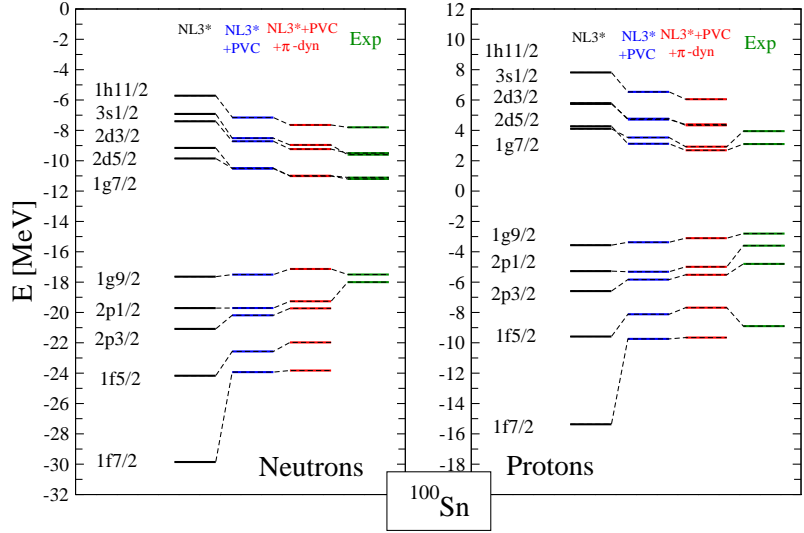}}
  \caption{Neutron (left) and proton (right) single-particle states in $^{100}$Sn. See text for explanations.}
 \label{100sn} 
\end{figure}
 As an illustration of the effect of the isovector phonons, single-particle energies of $^{100}$Sn were considered. The Dyson's equation for the single-particle propagator \cite{LR.06,L.12} was solved numerically with the self-energy of Fig. \ref{sfe}, however, neglecting the last terms represented by the backward going diagrams (they will be investigated in future work).  Isospin-flip vibrational states with spins and parities $J^{\pi} = 2^{\pm}, 3^{\pm}, 4^{\pm}, 5^{\pm}, 6^{\pm}$ and frequencies below 20 MeV were included into the self-energy. The results for {\it dominant} single-particle states are shown in Fig. \ref{100sn}. 
The mean-field states are displayed by the black bars, the PVC results with only isoscalar phonons are given by blue color, the PVC with both isoscalar and isovector phonons are marked with red, and the green bars represent 'experimental' single-particle energies extrapolated from data \cite{GBST.14,GLM.07}. One can see that the shifts of the dominant single-particle states due to the coupling to isovector phonons amount from few hundred keV to about 1 MeV. The spectroscopic factors are less sensitive and show only a small change compared to those obtained in Ref. \cite{LA.11}. 
Overall, it can be concluded that, except for a few cases, the inclusion of isovector phonons into the nucleonic self-energy improves the description of the single-particle states by shifting the dominant states in direction of the data. The origin of the remaining discrepancies and possibly still missing mechanisms will be studied in future work.
\section{SUMMARY}
Spin-isospin modes of excitation in neutron-rich nuclei were studied. It has been found that in some nuclei spin-isospin-flip states with angular momenta L$>$1 appear at very low energies with high transition probabilities. The two cases of the unstable nuclei $^{132}$Sn and $^{78}$Ni are investigated in detail within a relativistic approach. 
It is suggested that the observed collective spin-isospin-flip modes can strongly couple to other elementary modes, in particular, to single-particle ones. Such a coupling  has been introduced in the single-nucleon self-energy, in addition to the coupling to isoscalar phonons implemented in earlier studies.  In other words, the dynamical contribution of pion-exchange interaction has been taken into account on equal footing with the contribution of other mesons. The effect of this pion contribution on nuclear single-particle states is studied numerically for $^{100}$Sn. It has been shown that pion dynamics causes further redistribution of the single-particle strength while shifts of the dominant states amount up to 1 MeV and in most of the cases improve the agreement with data. Thus, a connection between the soft pionic modes and single-particle degrees of freedom is established and the shell structure is found to be sensitive to pion dynamics.

% Acknowledgement
\section{ACKNOWLEDGMENTS}
Fruitful discussions with E.E. Kolomeitsev, P. Ring, V. Tselyaev, T. Otsuka, and V. Zelevinsky are gratefully acknowledged. The author is very thankful to T. Marketin for providing the codes for pn-RRPA matrix elements. This work was supported by US-NSF grants PHY- 1204486 and PHY-1404343.
% References

%\nocite{*}
%\bibliographystyle{aipnum-cp}%
%\bibliography{nsd2015_litvinova}%

\end{document}